\documentclass{ws-ijmpcs}

\begin{document}

\markboth{Authors' Names}
{Instructions for Typing Manuscripts (Paper's Title)}

%
\catchline{}{}{}{}{}
%

\title{FORMATION AND COLLIMATION OF RELATIVISTIC MHD JETS - SIMULATIONS AND RADIO MAPS}

\author{CHRISTIAN FENDT}
\address{Max Planck Institute for Astronomy, K\"onigstuhl 17, \\
69117 Heidelberg, Germany,
fendt@mpia.de}

\author{OLIVER PORTH}
\address{Department of Applied Mathematics, The University of Leeds, \\
Leeds, LS2 9GT, United Kingdom\\
   }

\author{SOMAYEH SHEIKHNEZAMI}
\address{Max Planck Institute for Astronomy, K\"onigstuhl 17, \\
69117 Heidelberg, Germany
   }

\maketitle

\begin{history}
\received{Day Month Year}
\revised{Day Month Year}
\end{history}

\begin{abstract}
We present results of magnetohydrodynamic (MHD) simulations of jet formation and propagation,
discussing a variety of astrophysical setups.
In the first approach we consider simulations of relativistic MHD jet formation, considering jets
launched from the surface of a Keplerian disk, demonstrating numerically - for the first 
time - the self-collimating ability of relativistic MHD jets.
We obtain Lorentz factors up to $\simeq 10$ while acquiring a high degree of collimation
of about $1$ degree.
We then present synchrotron maps calculated from the intrinsic jet structure derived 
from the MHD jet formation simulation.
We finally present (non-relativistic) MHD simulations of jet lauching, treating the
transition between accretion and ejection. These setups include a physical magnetic
diffusivity which is essential for loading the accretion material onto the outflow.
We find relatively high mass fluxes in the outflow, of the order of 20-40\% of the
accretion rate. 
\end{abstract}

 \ccode{PACS numbers: 98.62.Mw, 95.30.Qd, 98.62.Nx, 98.38.Fs, 98.58.Fd}

\section{Introduction}
Astrophysical jets are highly collimated streams of high-velocity. 
They are observed in a variety of astronomical sources, such as young 
stars, micro-quasars, or active galactic nuclei (AGN).
The current understanding is that outflows are launched by 
{\em magnetohydrodynamic} (MHD) processes in the close vicinity of the
central object - an accretion disk surrounding a protostar or a compact object
\cite{1,2,3}.
Besides the contribution of the disk wind to the overall jet flow, there is 
most probably also a contribution launched by the central object.
In case of stellar sources, a stellar wind may contribute additional Poynting flux
or pressure \cite{4,5,6}.
In case of a central black hole, a central Poynting dominated spine jet may exist,
driven by the Blandford-Znajek mechanism \cite{7,8}.
Despite a huge effort, the details of all the physical 
processes involved are, however, not completely understood. 
Early treatments considered the stationary MHD equations, for example collimating
magnetic jets in Kerr metric \cite{9,10,11}, but relativistic jet simulations are
feasible since some decade \cite{12,13,14}.

Here we present results of axisymmetric MHD simulations
investigating certain aspects of jet formation.
We first present results for disk winds collimating into
relativistic jets applying (special) relativistic MHD simulations from
the surface of a Keplerian disk.
We then apply a relativistic radiation transfer code to provide synchrotron 
maps of these outflows - applying the true internal structure derived
by the numerical simulations.
We finally present simulations of jet launching. These simulations treat the
accretion and ejection process together, and therefore allow to derive
the ejection efficiency of the accretion disk.
As accretion is a (turbulently) diffusive process, these simulations
apply an $\alpha$-prescription of magnetic diffusivity. Since our code cannot
treat diffusivity in the relativistic limit, these simulations are
non-relativistic.
For all our simulations we have applied the MHD code PLUTO \cite{15}.

\section{Relativistic MHD jets from accretion disks}
Here we discuss (special) relativistic MHD simulations treating the formation 
of MHD jets from the surface of a disk.
With {\em formation} we denote the acceleration and collimation of a disk wind into
a high speed jet.
In order to apply a Keplerian disk boundary condition for the jet,
we have added Newtonian gravity to the special relativistic PLUTO code
(see \cite{16,17}).
Our standard setup involves a poloidal field considering a plasma-$\beta$ between
0.001 and 1 at the inner disk radius and the poloidal field profile as a fixed-in-time 
boundary condition.
The toroidal magnetic field is induced during the flow evolution and is floated into the
boundary domain.

\begin{figure}[t!]
\centering
\includegraphics[width=9cm]{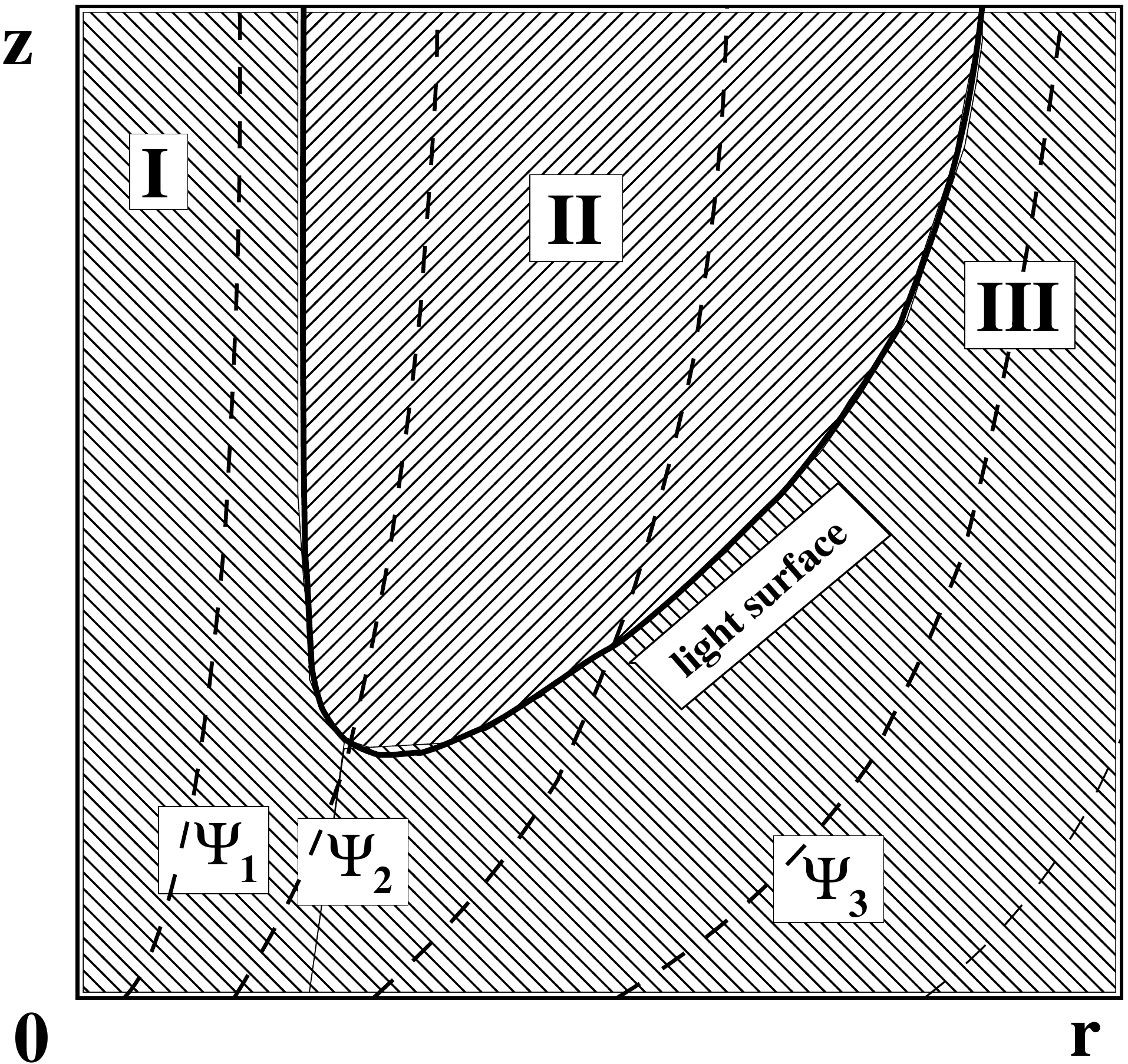}
\caption{The dynamical regimes of special relativistic
disk-winds. Region I and III stay sub-relativistic, region II
is relativistic, i.e. electric forces are not negligible.
Figure taken from [16].
}
\label{fig:lightcylinder}
\end{figure}

Figure \ref{fig:lightcylinder} shows the relativistically distinct regimes of flow collimation.
In the hydrodynamic regime upstream the Alfv\'en surface (I),
gravity balances thermal and magnetic pressure, respectively the centrifugal force for 
cold jets. 
In the relativistic regime downstream the light surface (II), the poloidal magnetic
pressure gradient imposes a collimating force against electric de-collimation.
Electric forces ultimately overcome the classical magneto-centrifugal contribution.
In the magneto-hydrodynamic regime downstream the Alfv\'en surface (III)
the residuals of magnetic pinch and toroidal magnetic pressure gradient balance
the centrifugal force \cite{16}.

\begin{figure*}[t!]
 \begin{center}
 \includegraphics[height=9.0cm]{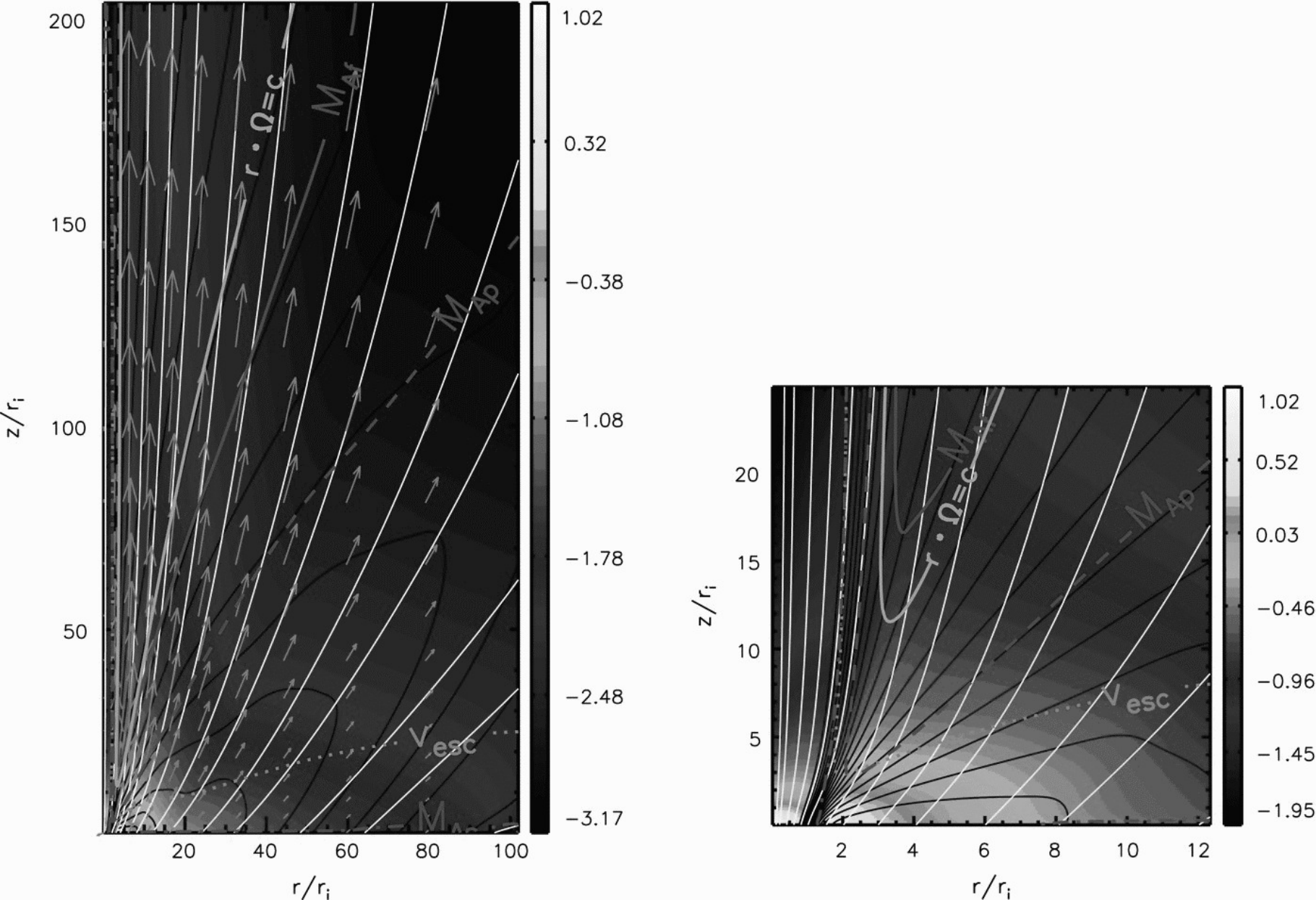}
 \end{center}
 \caption{
Rest-frame density (grey colors, logarithmic) of the stationary flow.
Shown are poloidal magnetic field lines (solid white),
electric stream lines (solid black),
characteristic MHD surfaces (various dot-dashed grey, labeled
as $M_{\rm As}$, $M_{\rm Ap}$, $M_{\rm Af}$),
the surface of escape velocity (dotted grey, labeled as $V_{\rm esc}$),
the light surface (solid grey, labeled as $r\cdot \Omega = c$).
Arrows indicate the velocity field.
The right figure is the enlarged central region,
indicating the three regimes defined by the light surface.
Figures taken from [16].
    }
\label{fig:relmhd}
\end{figure*}

Figure \ref{fig:relmhd} shows the typical result of a mildly relativistic jet
with Lorentz factor of 1.5. The location of the magnetosonic surfaces is shown
and also the light surface where $R_{\rm L}(r,z) \equiv c/\Omega$.
The asymptotic region of the relativistic jet enclosed by the light surface is
the {\em truly relativistic part of the outflow}.
This part of the outflow originates close to the inner edge of the accretion disk
deep in the potential well where rotation is most rapid.
We have also applied a model setup where we prescribe a higher Poynting flux by 
injecting plasma with a higher toroidal field strength (up to a factor 8) from 
the disk into the outflow.
Depending on the Poynting flux injected, the resulting jet Lorentz factors
reach number values between of 1.5 - 10.
Typical jet opening angles for the relativistic part of the outflow are $1-7$ degrees,
however, on the physical scales covered by the simulations of up to $6000^2$
Schwarzschild radii the asymptotic jet is not jet fully accelerated or collimated.

We interpret our results as a numerical proof of MHD jet self-collimation.
We have applied a modified outflow boundary condition, which is force-free
concerning the Loretz forces.
Moreover, the poloidal magnetic field strength as well as the gas pressure
decrease with axial distance from the jet axis.
Therefore, no collimating forces are present from these terms,
but only the tension force of the toroidal field.
This is the first time that the predictions by
Heavearts \& Norman \cite{18} and Chiueh et al. \cite{19} of MHD jet self-collimation
are proven by numerical simulations for relativistic jets.
Note that we particularly calculate the force-equilibrium across the light surface
and do {\em not} apply the conventional limit for asymptotic approximations of
$r >> r_{\rm L}$. Further, there is no outer medium present which collimates the outflow.

\section{Synchrotron maps of simulated MHD jets}
\begin{figure*}[t!]
 \begin{center}
 \includegraphics[height=6.0cm]{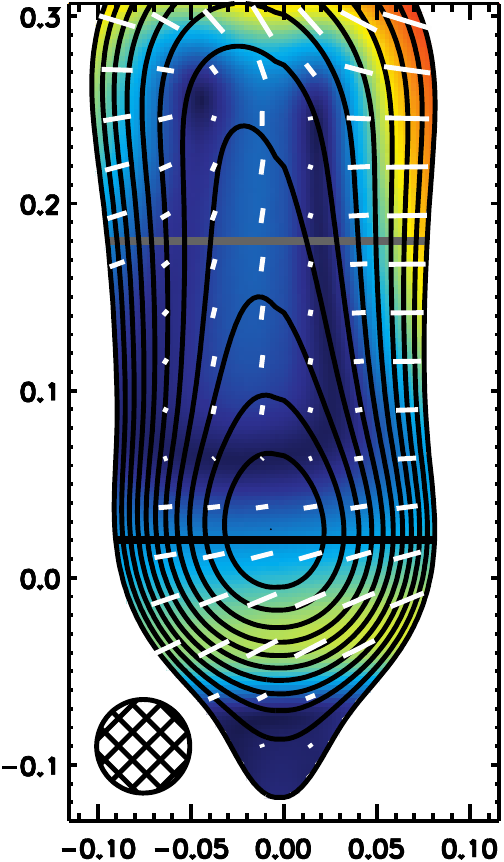}
 \includegraphics[height=6.0cm]{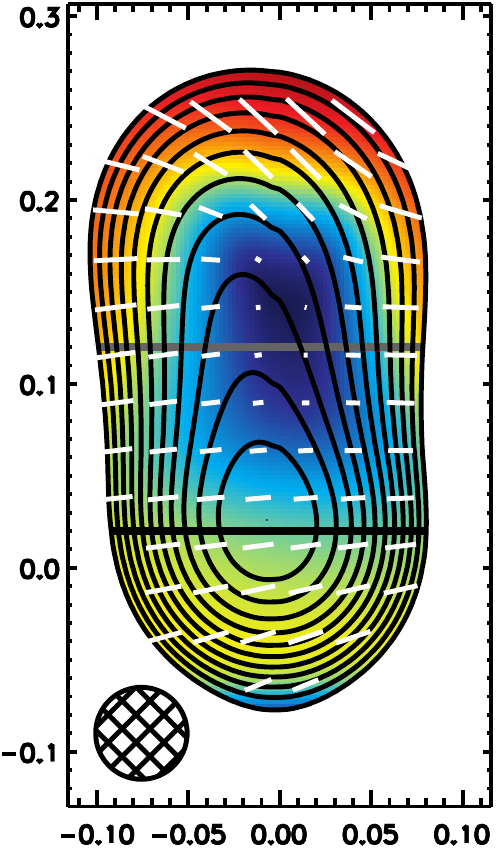}
 \includegraphics[height=6.0cm]{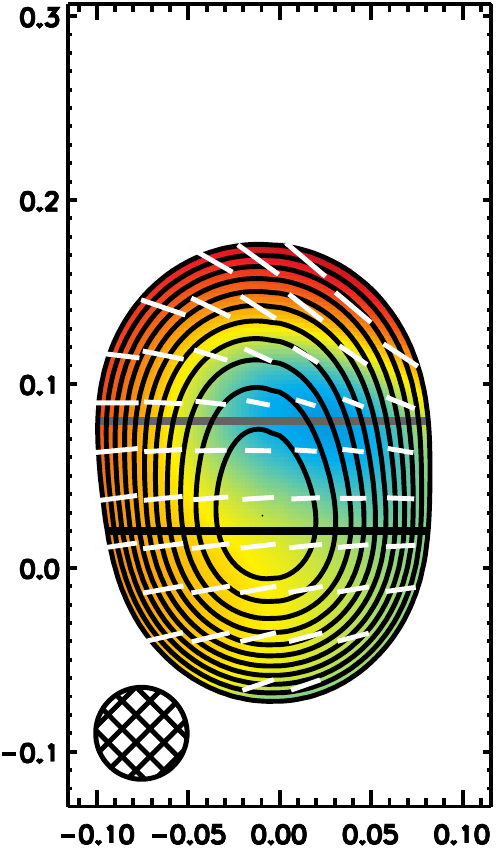}
 \includegraphics[height=6.0cm]{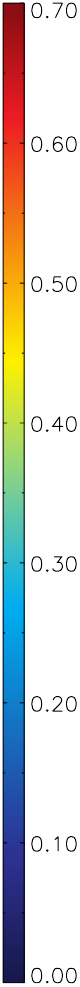}

 \includegraphics[height=6.0cm]{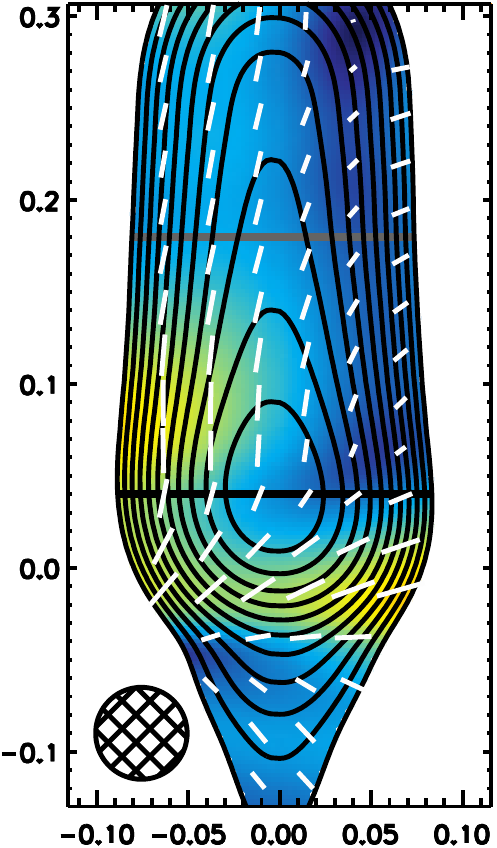}
 \includegraphics[height=6.0cm]{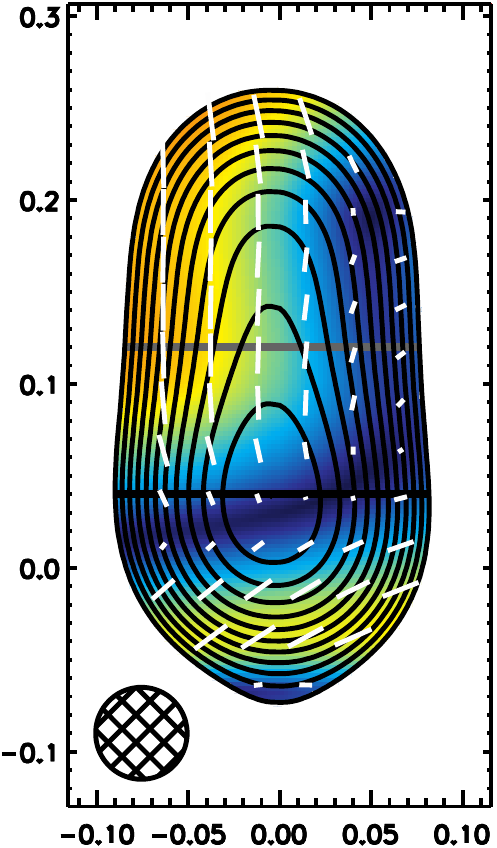}
 \includegraphics[height=6.0cm]{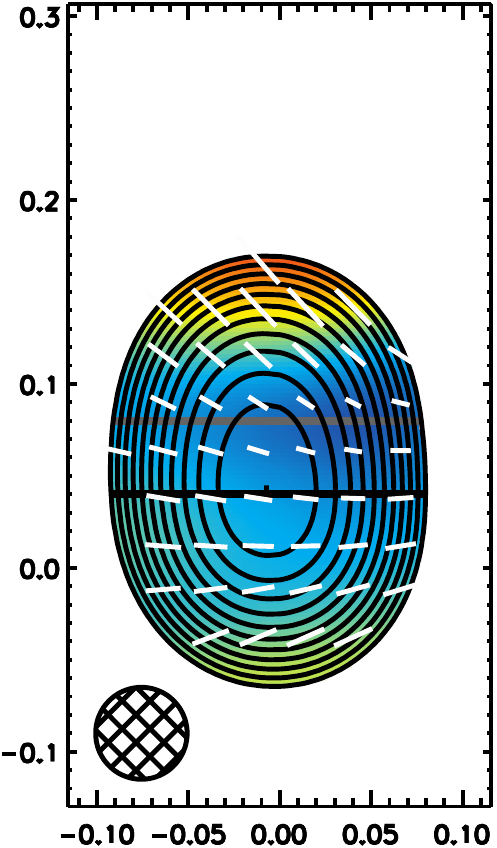}
 \includegraphics[height=6.0cm]{fendt_f3g.pdf}

 \end{center}
 \caption{
 Polarizations for inclination 30, 20, 10 degree (left to right) emitted from regions 
 with comoving pitches $B'_{\phi} / B'_{\rm p} > 1$ (above) 
 and $B'_{\phi} / B'_{\rm p} > 2$ (below). 
 The polarization degree $\Pi$43 GHz is color-coded, and intensity 43 GHz contours 
 are shown. 
 The spatial scale is given in milliarcseconds, and a restoring beam with FWHM = 0.05 
 mas was used. Figures taken from [17].
    }
\label{fig:radio}
\end{figure*}

Having derived a physically consistent distribution of the MHD variables
by our numerical simulations - i.e. the jet magnetic field distribution, the density distribution,
and the jet velocity - we may us this information to derive consistent radio synchrotron maps
following a relativistic polarized radiation transfer within the jet. 
Since we have not included particle acceleration in the MHD model, additional assumptions have
to be made.
We have tested three tracers for the power-law particle acceleration, i.e. density, thermal
pressure, or magnetic energy density \cite{17}.
All tracers give a similar polarization structure, although the intensity distribution
differs.
Figure \ref{fig:radio} shows
the polarization degree and polarisation vectors for jet inclination 30, 20, and 10 degrees,
emitted from regions with different comoving pitches $B'_{\phi} / B'_{\rm p}$.

\section{Jet launching from accretion disks}
We now show simulations treating the {\em launching} of the outflow - i.e. 
the transition from accretion to ejection.
Accretion disk physics usually relies on (turbulent) magnetic diffusivity
(unless a high-enough resolution is applied resolving the 
magneto-rotational instability).
Only few relativistic codes are able to apply a physical magnetic 
resistivity (see e.g.~[20]). In order to treat the
diffusive MHD launching problem, we have therefore applied a non-relativistic
approach \cite{21,22}.

\begin{figure*}[t!]
 \begin{center}
 \includegraphics[height=8.0cm]{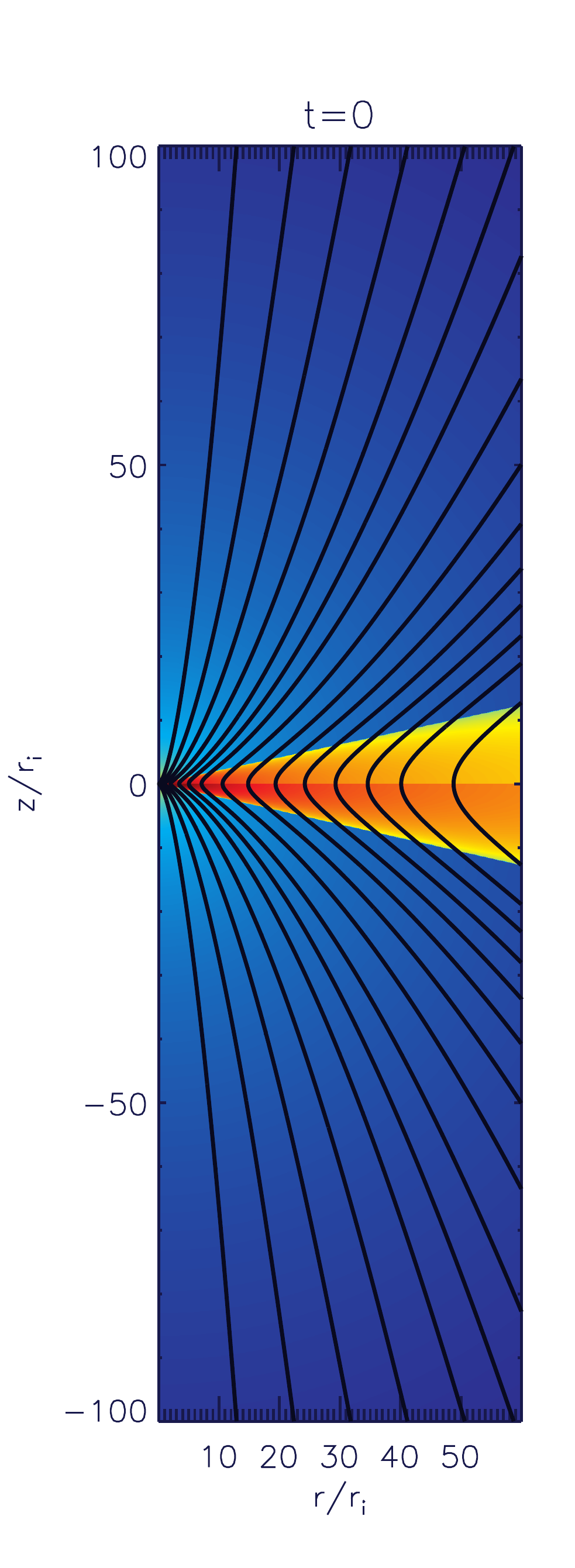}
 \includegraphics[height=8.0cm]{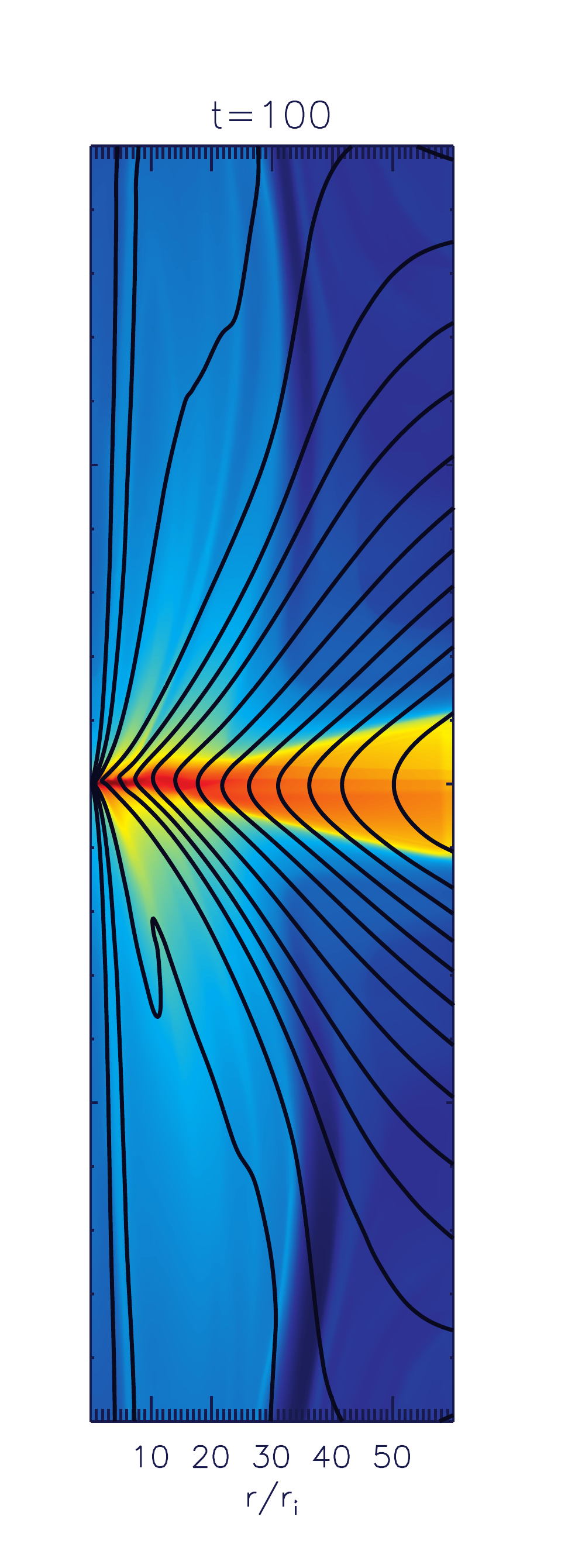}
 \includegraphics[height=8.0cm]{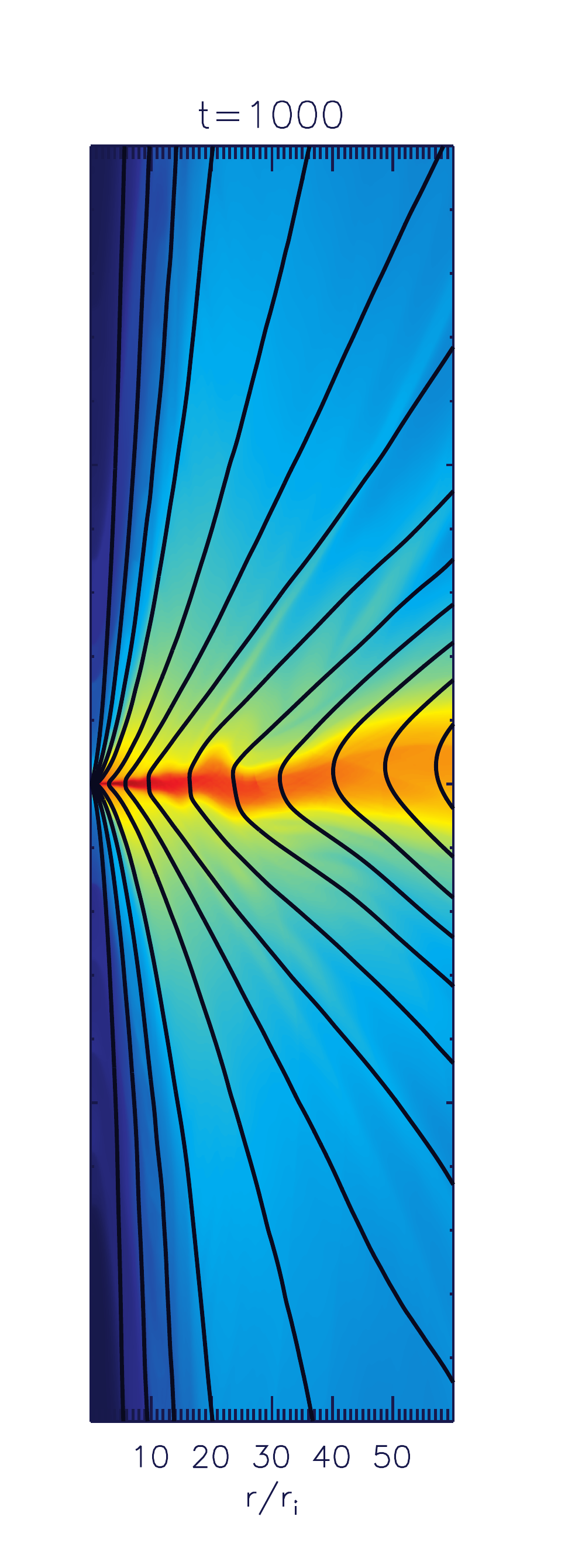}
 \includegraphics[height=8.0cm]{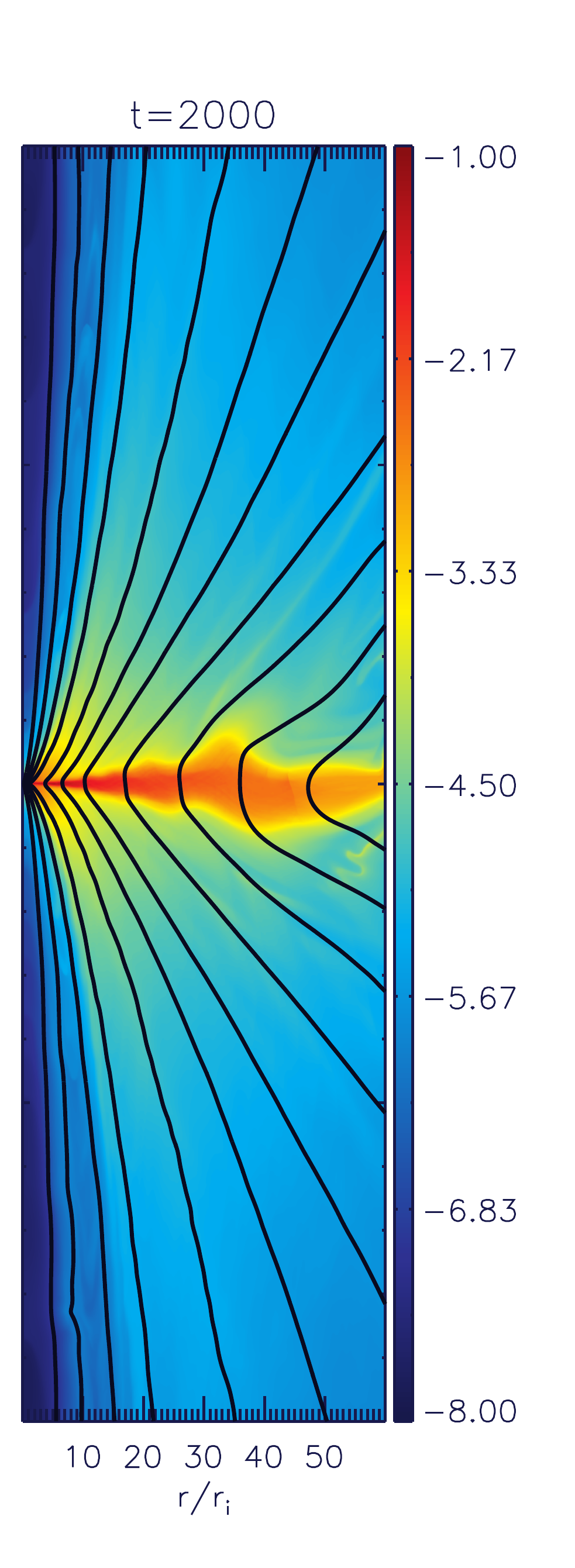}
 \end{center}
 \caption{
Time evolution of the bipolar jet-disk structure for a simulation starting from an
initial initial state with different thermal scale heights for the upper and lower 
disk hemispheres. Shown is the evolution for dynamical time t = 0, 100, 1000, 2000 
of the mass density (colors) and the poloidal magnetic field (lines).
Figures taken from [22].
    }
\label{fig:launching}
\end{figure*}

Figure \ref{fig:launching} shows results from our bipolar simulations. The simulations were
run from a symmetric initial condition, apart from a hemispheric asymmetry 
in the disk pressure distribution. This disk asymmetry leads first to disk
warping and, subsequentely, to asymmetric jet and counter jets differing
in mass flux or velocity by about 30\%. Depending on the magnetic diffusivity
prescription (global model or local model), the jet - counter jet asymmetries
are long lasting. We also observed a reversal of the asymmetry, meaning that
the jet with high mass flux weakens and becomes less massive than the
counter jet. The time scale for these reversals is about 1000 dynamical
time scales of the inner disk.
For AGN jets, the timescale for jet fluctuations corresponds to 2 yr, assuming 
an inner disk radius of 10 Schwarzschild radii and a central black hole mass of
$10^8$ solar masses. 
This is remarkably similar to the ejection times observed in 
e.g. 3C 120 \cite{23} or 3C 390.3 \cite{24}.

\section*{Acknowledgments}
This work was partly financed by the SFB 881 of DFG, project B4, 
and the IMPRS for Astronomy \& Cosmic Physics
at the University of Heidelberg. We thank Andrea Mignone for the possibility to use 
the PLUTO code. 
The simulations were performed on the THEO cluster of the Max Planck Institute for Astronomy. 
\appendix


\end{document}